\documentclass[amsfonts, prd, twocolumn, nofotinbib, showpacs]{revtex4}
\usepackage{amssymb}
\usepackage{amsmath}
\usepackage{amsfonts}
\usepackage{graphicx, float}
\usepackage{graphicx, epsfig}
\usepackage{color}
\usepackage{enumerate}
\newcommand{\beq}{\begin{equation}}
\newcommand{\eeq}{\end{equation}}
\newcommand{\bea}{\begin{eqnarray}}
\newcommand{\eea}{\end{eqnarray}}

\newcommand{\Mpl}{M_{\rm Pl}}
\newcommand{\Madm}{M_{\rm ADM}}
\newcommand{\ellp}{\ell_{\rm Pl}}

\newcommand{\RC}{R_{\rm C}}
\newcommand{\RS}{R_{\rm S}}
\newcommand{\Tpl}{T_{\rm Pl}}
\newcommand{\tpl}{t_{\rm Pl}}
\newcommand{\rH}{r_{\rm H}}

\begin{document}
\title{Sub-Planckian black holes and the Generalized Uncertainty Principle}

\author{Bernard Carr$^1$\footnote{E-mail: b.j.carr@qmul.ac.uk}, Jonas Mureika$^2$\footnote{E-mail: jmureika@lmu.edu} and Piero Nicolini$^{3,4}\footnote{E-mail: nicolini@fias.uni-frankfurt.de}$}
\affiliation{$^1$Astronomy Unit, Queen Mary University of London, Mile End Road, London E1 4NS, UK\\
$^2$Department of Physics, Loyola Marymount University, 1 LMU Drive, Los Angeles, CA, USA~~90045\\
$^3$Frankfurt Institute for Advandced Studies (FIAS), 
Ruth-Moufang-Stra$\beta$e 1, 1,60438 Frankfurt am Main, Germany\\
$^4$Institut f\"ur Theoretische Physik, Johann Wolfgang 
Goethe-Universit\"at, Max-von-Laue-Stra$\beta$e 1, 60438 Frankfurt am Main, Germany}


\begin{abstract}
The Black Hole Uncertainty Principle correspondence suggests that there could exist black holes with mass beneath the Planck scale but radius of order the Compton scale rather than Schwarzschild scale.  We present a modified, self-dual Schwarzschild-like metric that reproduces desirable aspects of a variety of disparate models in the sub-Planckian limit, while remaining Schwarzschild in the large mass limit. The self-dual nature of this solution under $M \leftrightarrow M^{-1}$ naturally implies a Generalized Uncertainty Principle with the linear form $\Delta x \sim \frac{1}{\Delta p} + \Delta p$.   We also demonstrate a natural dimensional reduction feature, in that the gravitational radius and thermodynamics of sub-Planckian objects resemble that of $(1+1)$-D gravity. The temperature of sub-Planckian black holes scales as $M$ rather than $M^{-1}$ but the evaporation of those smaller than $10^{-36}$g is suppressed by the cosmic background radiation.
This suggests that relics of this mass could provide the dark matter.
\end{abstract}



\maketitle

\section{Introduction}
One of the foundational tenets of the microscopic domain is the
Heisenberg Uncertainty Principle (HUP) which implies that the uncertainty in the position and momentum of a particle must satisfy
$ \Delta x > \hbar/ (2 \Delta p)$.
An important role is therefore played by the reduced Compton wavelength,
$\RC = \hbar/(Mc)$, 
which can be obtained from the HUP with the substitution $\Delta x \rightarrow R$ and $\Delta p \rightarrow c M$ but without the factor of $2$. \begin{figure}[b]
\includegraphics[scale=.35]{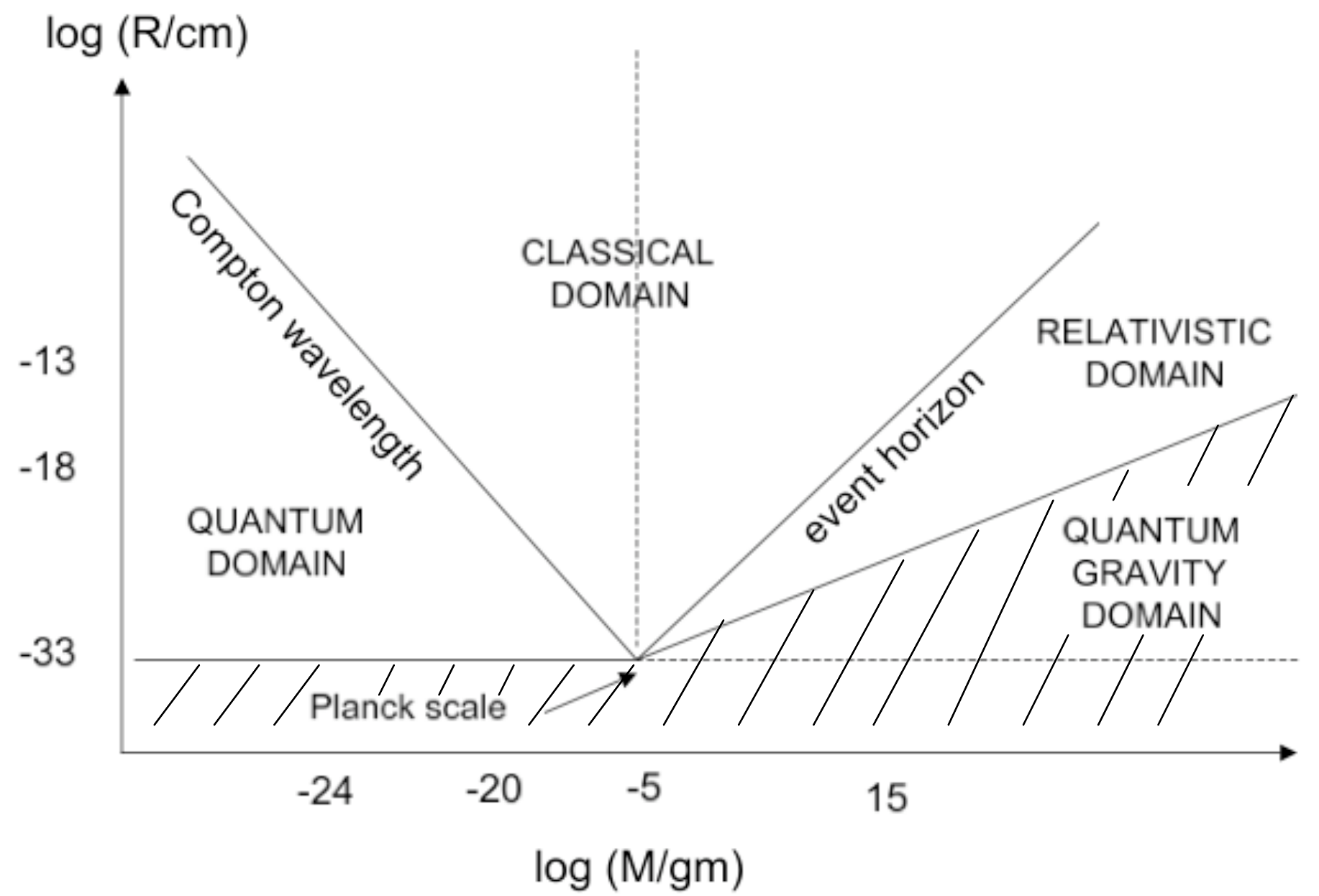}
\caption{The division of the ($M,R$) diagram into the classical, quantum, relativistic and quantum gravity domains.} 
\label{MR}       
\end{figure}
Similarly, a key feature of the macroscopic domain is the existence of black holes,  
 a spherically symmetric object of mass $M$ forming an event horizon if it falls within its Schwarzschild radius, $\RS = 2GM/c^2$.

As indicated in Fig.~\ref{MR}, the Compton and Schwarzschild lines
 intersect at around the Planck scales,
$ \ellp = \sqrt{ \hbar G/c^3} \sim 10^{-33} \mathrm {cm}, 
 \Mpl = \sqrt{ \hbar c/G} \sim 10^{-5} \mathrm g $.
Here the vertical line 
$M=\Mpl$ is often assumed to mark the division between  elementary particles ($M <\Mpl$) and black holes ($M > \Mpl$), because one usually requires a black hole to be larger than its own Compton wavelength.
The horizontal line $R=\ell_{\rm Pl}$ 
is significant because quantum fluctuations in the metric should  become important below this.
Quantum gravity effects should also be important whenever the density exceeds the Planck value,
{\it i.e.} when $R <( M/\Mpl)^{1/3}\ellp $,
which is well above the $R = \ellp$ line in Fig.~\ref{MR} for $M \gg \Mpl$.

\begin{figure}[b]
\includegraphics[scale=.4]{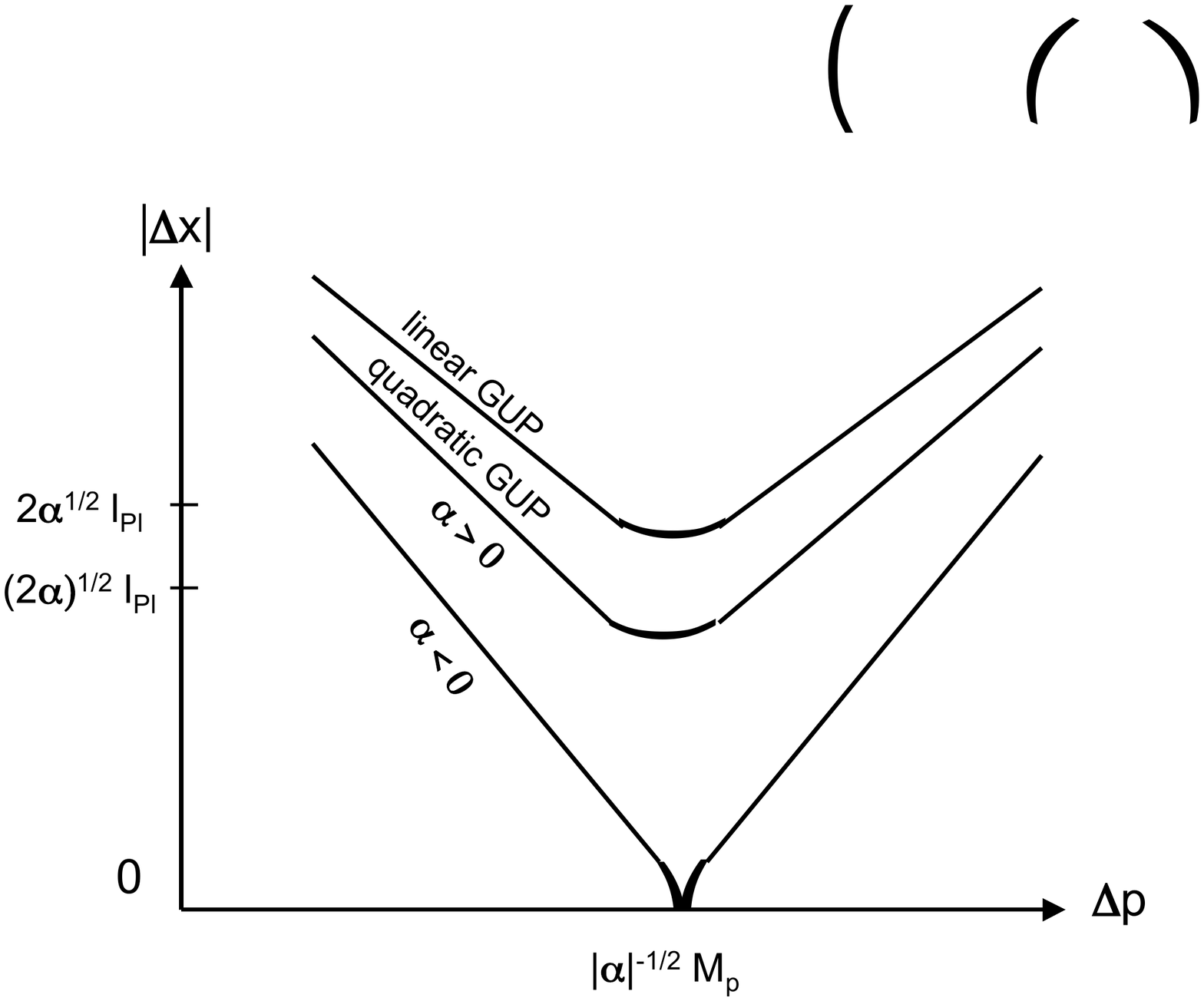}
\caption{The upper curves show $|\Delta x|$ versus $\Delta p$ for the GUP with $\alpha > 0$ in its linear and quadratic forms,  these giving a smooth transition between particles and black holes.  For $\alpha < 0$, $|\Delta x|$ has a minimum at $0$ with a discontinuus gradient, 
so that the black hole and particle states are disconnected. For more general forms of the  GUP and GEH, in which the parameters $\alpha$ and $\beta$ are independent, the relationship between particles and black holes is more complicated.  }
\label{modesto1}       
\end{figure}

Although the Compton and Schwarzschild boundaries 
correspond to straight lines in the logarithmic plot of Fig.~\ref{MR}, this form presumably breaks down near the Planck point. As one approaches the Planck point from the left in Fig.~\ref{MR}, it has been argued~\cite{Adler_1, Adler_2, Adler_3, Adler_4} 
that the HUP should be replaced by a Generalized Uncertainty Principle (GUP) of the form
\begin{equation}
\Delta x > \frac{ \hbar}{\Delta p }+ \left( \frac{ \alpha \ellp^2 }{ \hbar} \right) \Delta p \, .
\label{GUP1} 
\end{equation}
Here $\alpha$ is a dimensionless constant (usually assumed positive) which depends on the particular model and the factor of $2$ in the first term has been dropped . 
This form of the GUP is indicated by the top curve in Fig.~\ref{modesto1}.
Variants of (\ref{GUP1}) can be found in other approaches to quantum gravity, including
loop quantum gravity~\cite{ashtekar_1,ashtekar_2}, string theory~\cite{veneziano_1, veneziano_2, veneziano_3, veneziano_4, veneziano_5, veneziano_6}, non-commutative quantum mechanics \cite{majid}, gravity ultraviolet self-completeness \cite{nicolini} and
general minimum length considerations \cite{maggiore_1,maggiore_2,maggiore_3}.

If we rewrite (\ref{GUP1}) using the same substitution $\Delta x \rightarrow R$ and $\Delta p \rightarrow c M$ as before, 
it becomes
\beq
R > \RC' \equiv \frac{ \hbar}{Mc} +\frac { \alpha GM}{c^2} = \frac{\hbar}{Mc} \left[ 1  + \alpha  \left( \frac {M}{\Mpl}\right)^2 \right] \, .
\label{GUP2}
\eeq
This expression might be regarded as a generalized Compton wavelength,
the last term representing a small correction as one approaches the Planck point from the left.
However, one can also apply (\ref{GUP2}) for $M \gg \Mpl$ and it is interesting that in this regime it asymptotes to the Schwarzschild 
form, apart from a numerical factor \cite{carr_1}. This suggests that  there is a different kind of positional uncertainty for an object larger than the Planck mass, related to the existence of black holes. 
This is not unreasonable since the Compton wavelength is below the Planck scale (and hence meaningless) here. Also an outside observer cannot localize an object on a scale smaller than its Schwarzschild radius. 

The GUP has important implications for the black hole horizon size, as can be seen by  examining what happens as one approaches the intersect point from the right Fig.~\ref{MR}. In this limit, 
it is natural to write (\ref{GUP2}) as
 \begin{equation}
R > \RS' = \frac{\alpha GM}{c^2} \left[ 1  + \frac{1}{\alpha} \left( \frac {\Mpl}{M}\right) ^2 \right] 
\label{GEH2}
\end{equation}
and this represents a small perturbation to the Schwarzschild radius for $M \gg \Mpl$ 
if one assumes $\alpha =2$. 
There is no reason for anticipating $\alpha=2$ in the heuristic derivation of the GUP. 
However, the factor of 2 in the expression for the Schwarzschild radius is precise, whereas  the coefficient associated with the Compton term is somewhat arbitrary. This motivates an alternative approach in which the free constant in (\ref{GUP2}) is associated with the first term rather than the second. 
One then replaces Eqs.~(\ref{GUP2}) and (\ref{GEH2}) with the expressions
 \begin{equation}
\RC' = 
\frac{\beta \hbar}{Mc} \left[ 1  + \frac{2}{\beta} \left( \frac {M}{\Mpl}\right)^2 \right] 
\label{GUP4A}
 \end{equation}
and 
\begin{equation}
\RS' = \frac{2 GM}{c^2} \left[ 1  + \frac{\beta}{2} \left( \frac {\Mpl}{M}\right) ^2 \right] 
\label{GUP4B}
\end{equation}
for some constant $\beta$, with
the second expression being regarded as a Generalized Event Horizon (GEH). 

An important caveat is that
(\ref{GUP1}) assumes the two uncertainties add linearly.
 On the other hand, 
since they are independent, it might be more natural to assume that they add quadratically \cite{cmp}:
 \begin{equation}
  \Delta x > \sqrt {\left( \frac{\hbar}{ \Delta p}\right)^2 + \left( \frac{ \alpha \ellp^2 \Delta p}{ \hbar} \right)^2} \, .\label{quad2}
 \end{equation}
This corresponds to the lower GUP curve in Fig.~\ref{modesto1}. 
We refer to Eqs.~(\ref{GUP1}) and (\ref{quad2}) as the {\it linear} and {\it quadratic} forms of the GUP. 
Adopting the $\beta$ formalism, then gives a unified expression for  generalized Compton wavelength and  event horizon size 
\begin{equation}
\RC' = \RS' = \sqrt{ \left( \frac{\beta \hbar }{Mc}\right)^2  + \left( \frac {2GM}{c^2}\right)^2 } \, ,
\label{quadEH}
\end{equation}
leading to the approximations
 \begin{equation}
\RC' \approx \frac{\beta \hbar}{Mc}  \left[ 1  + \frac{2}{\beta^2} \left( \frac {M}{\Mpl}\right)^4 \right] 
\end{equation}
and
\begin{equation}
\RS' \approx \frac{2GM}{c^2} \left[ 1  + \frac{\beta^2}{8} \left( \frac {\Mpl}{M}\right) ^4 \right] 
\label{quadC}
\end{equation}
for $M \ll \Mpl$ and $M \gg \Mpl$, respectively. These might be compared to the {\it exact} expressions in the linear case, given by Eqs.~(\ref{GUP4A}) and (\ref{GUP4B}).

Regardless of the exact form of the GUP, these arguments suggest that there is a connection between the Uncertainty Principle on microscopic scales and black holes on macroscopic scales. This is termed the Black Hole Uncertainty Principle (BHUP) correspondence and it is manifested in a unified expression for the Compton wavelength and Schwarzschild radius \cite{carr_2}. 
It is a natural consequence of combining the notions of the GUP and the GEH. Indeed, it would be satisfied for any form of the function $\RC' \equiv \RS'$ which asymptotes to $\RC$ for $M \ll \Mpl$ and $\RS$ for $M \gg \Mpl$. Models in which this function is symmetric under the duality transformation $M \leftrightarrow 1/M$ (such as the linear and quadratic forms given above) are said to satisfy the {\it strong} BHUP correspondence \cite{carr_2}.

One controversial implication of the BHUP correspondence is that it suggests there could be sub-Planckian black holes with a size of order their Compton wavelength.  For example, one can argue that there is only a low probability of sub-Planckian objects becoming black holes \cite{casadio_1, casadio_2}.  
In fact, it has been claimed that loop quantum gravity already predicts the existence of such black holes  \cite{poly_1,poly_2,poly_3,poly_4,poly_6, poly_7}, with their radius having precisely the form (\ref{quadEH}) associated with the quadratic GUP \cite{cmp}. However, these `loop black hole' (LBH) solutions are really wormholes and they involve another asymptotic space. Their physical validity is contentious and there are also some technical subtleties associated with them. 
In this paper, we explore another type of solution which involves sub-Planckian black holes but avoids some of the complications associated with the LBH solution. In particular, it implies a linear rather than quadratic form of the GUP and it does not involve another asymptotic space.

The continuity between the Compton and Schwarzschild lines shown by the upper curves in  Fig.~\ref{modesto1} suggests some link between elementary particles and sub-Planckian black holes.
However, one might prefer to maintain a distinction between these objects. For example, 
the function  $|\Delta x|$ has a minimum at $0$ for models with $\alpha < 0$ (bottom curve) but with a discontinuity in the gradient. Since $R_C' = R_S' = 0$ at this point, one effectively has $G \rightarrow 0$  (no gravity) and  $\hbar \rightarrow 0$ (no quantum discreteness), which relates to models involving asymptotic safety \cite{bonnano} and world crystals \cite{scardigli}. 
The distinction between particles and black holes could also be maintained with more general forms of the GUP and GEH. For example, Sec.~II  considers the possibility that the parameter $\alpha$ describing the GUP and the parameter $\beta$ describing the GEH are independent. In this case, there are still sub-Planckian black holes but the relationship between these and elementary particles becomes more complicated. 

The plan of this paper is a follows. In Sec.~II we discuss the concept of mass using the Komar integral and find that this provides a useful way of linking black holes and elementary particles. Our definition of mass naturally implies a linear form for the GUP and it  also suggests that gravity is effectively 2-dimensional near the Planck scale. In Sec.~III we introduce a new black hole solution, which resembles Schwarzschild except that the mass is interpreted differently and relates to the definition given in Sec.~II.  We examine the thermodynamical properties of these solutions in Sec.~IV, using two different derivations of the temperature. These agree asymptotically in the large and small mass regimes and suggest  that the temperature of a sub-Planckian black holes is proportional to its mass. In Sec.~V we examine the cosmological consequences of this. We draw some general conclusions in Sec.~VI.

\section{Planck black holes and beyond}

Let us recall some basic ideas and open problems concerning black holes at the Planck scale.
In the standard picture, the Schwarzschild solution is obtained by solving Einstein's equations in vacuum and matching the metric coefficients with the Newtonian potential  as a boundary condition  to fix the integration constant. This constant relates to the mass specified by the Komar integral \citep[p. 251]{carroll}:
\begin{equation}
M\equiv\frac{1}{4\pi G}\int_{\partial\Sigma}d^2x\sqrt{\gamma^{(2)}}\ n_\mu\sigma_\nu\nabla^\mu K^\nu
\label{komar}
\end{equation}
where $K^\nu$ is a timelike vector, $\Sigma$ is a spacelike surface with unit normal $n^\mu$, and $\partial\Sigma$ is the boundary of $\Sigma$ (typically a 2-sphere at spatial infinity) with metric $\gamma^{(2)ij}$ and outward normal $\sigma^\mu$. 

For a large mass black hole, $M\gg\Mpl$, quantum effects are negligible and one finds the usual Schwarzschild solution. For a sub-Planckian object ($M<\Mpl$), however, the mass parameter can simultaneously refer to a particle and a Schwarzschild black hole. 
Usually one rejects the existence of such black holes because $\RS$ would be smaller that $\ellp$, making the application of general relativity unreliable.   As a result,  one generally considers only the particle case in the sub-Planckian regime and writes \eqref{komar} as   
\begin{equation}
M\equiv \int_{\Sigma} d^3x \sqrt{\gamma}\ n_\mu K_\nu T^{\mu\nu}\simeq -4\pi\int_0^{\RC} dr \, r^2  T^{\ 0}_{ 0}
\label{komarparticle}
\end{equation}
where $\gamma$ is the determinant of the spatially induced metric $\gamma^{ij}$, $T^{\mu \nu}$ is the stress-energy tensor and $T^{\ 0}_{ 0}$ accounts for the particle distribution on a scale of order $\RC$. This corresponds to the rest mass appearing in the expression for the Compton wavelength, $\RC=\hbar/(Mc)$.

Let us now consider another situation: a decaying black hole with mass $M\gtrsim \Mpl$. The fate of such an object is an open problem in quantum gravity and connects with the definition of the mass.
There are at least three possible scenarios for the end-point of evaporation. 

(i) 
The black hole keeps decaying semi-classically with a runaway increase of the temperature and a final explosion involving  non-thermal emission of hard quanta. In this case, the energy momentum tensor exhibits an 
integrable singularity, $T^{\ 0}_{ 0} = -M\delta(\vec{x})$,
and the Komar energy has a standard profile. However, this scenario may be criticized since it relies on classical and semi-classical arguments applied to a quantum gravity dominated regime. 

(ii) 
Quantum gravity effects modify the classical profile of the mass-energy distribution, so that $T^{\ 0}_{ 0}\neq -M\delta(\vec{x})$. This happens in a variety of proposals, including asymptotically safe gravity \cite{alfio}, non-commutative geometry \cite{piero, piero2, pnes10}, non-local gravity \cite{nicolini, mmn11, piero3} and gravitational self-completeness \cite{dvali,  piero4, antonio1, antonio2}. In all these cases, the end-point of evaporation turns out to be a stable zero-temperature extremal black hole configuration, preceded by a positive heat capacity cooling 
phase. The Komar energy would again be defined by \eqref{komar}, while the size of the black hole remnant would correspond to the natural ultraviolet cut-off of quantum gravity.  This means that the endpoint of evaporation separates the two phases, {\it i.e.,}  particles and black holes.
Such a scenario has the following three properties: 
\begin{itemize}
\itemsep1pt \parskip0pt \parsep0pt
\item singularity avoidance or inaccessibility
\item non-singular final stage of evaporation
\item consistent definition of black hole size with $\RS>\ellp$ for all masses.
\end{itemize}
Despite the virtues of such a proposal, we do not yet have a universally recognized principle to support it or any proof that it represents the unique alternative to the semiclassical scenario (see \cite{sabine} for a further discussion). Only a self-consistent theory of quantum gravity can confirm this possibility.

(iii) In the absence of further theoretical indications or experimental evidence, we explore a third scenario, which reverses the usual logic but still assumes the above three properties. In so far as the black hole undergoes a final stage of evaporation, 
the major contribution to integral \eqref{komar} will be 
\begin{equation}
M = -4\pi\int_0^{\ellp} dr \, r^2  T^{\ 0}_{ 0}
\label{qbhkomar}
\end{equation}
where $T^{\ 0}_{ 0}$ accounts for an unspecified quantum-mechanical distribution of matter and energy. 
One still has $M \neq -M\delta(\vec{x})$ but the profile differs from the second scenario. Integral \eqref{qbhkomar} is generally  not known and might lead to a completely different definition of the Komar energy. Some anomalies are expected to emerge at the Planck scale since they already emerge at the GUP level. 

\subsection{Black Holes Versus Elementary Particles}

The parameter $M$ is simultaneously the particle rest mass and Komar energy in \eqref{GUP2} and \eqref{GUP4B}. 
Inspired by the dual role of $M$ in the GUP, we now explore a variant of the third 
scenario above, based on the existence of sub-Planckian black holes, \textit{i.e.} quantum mechanical objects  that are simultaneously black holes and  elementary particles; we dub these ``black particles''. 
In this context, we suggest that the Arnowitt-Deser-Misner (ADM) mass, which coincides with the Komar energy in the stationary case, should be 
\begin{equation}
\Madm = M \left(1+\frac{\beta}{2} \frac{\Mpl^2}{M^2}\right)
\label{newmass}
\end{equation}
since the associated value of $2GM/c^2$ then interpolates between the standard super-Planckian ($M\gg\Mpl$), trans-Planckian ($M \gtrsim \Mpl)$
and particle-like 
sub-Planckian ($M<\Mpl$) behaviours. One justification for \eqref{newmass} could be that this {\it implies} the GUP, since  
\beq
\frac{2\Madm}{r} = \frac{2 \Delta p}{c\Delta x} \, ,
\label{ADM}
\eeq
provided the momentum term is replaced by
\beq
\Delta p \longrightarrow \Delta p + \frac{\Mpl^2 c^2 }{\Delta p} \, .
\eeq
\if
The approach of a mass redefinition
\begin{equation}
M \rightarrow M \left(1+\frac{\beta}{2} \frac{\Mpl^2}{M^2}\right)
\label{newmass}
\end{equation}
 is inspired by the GUP duality.  Alternatively, the Adler approach assumes {\it a priori} the GUP.  
\fi
Given that the exact nature of gravitation at the Planck scale is unknown, there is perhaps no way to assess the plausibility of this argument.

\begin{center}
\begin{figure}[h]
\includegraphics[scale=0.35]{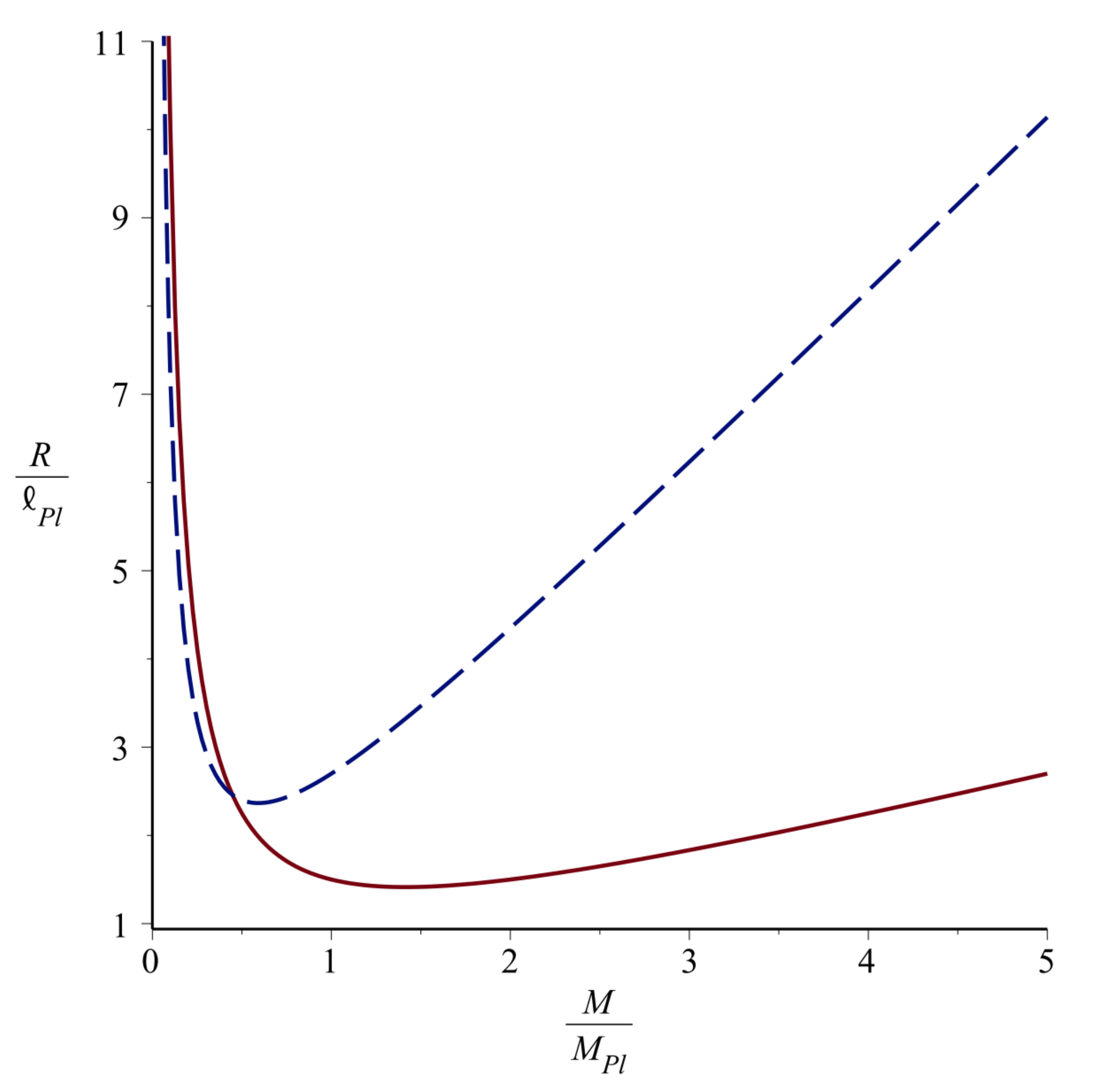}
\includegraphics[scale=0.35]{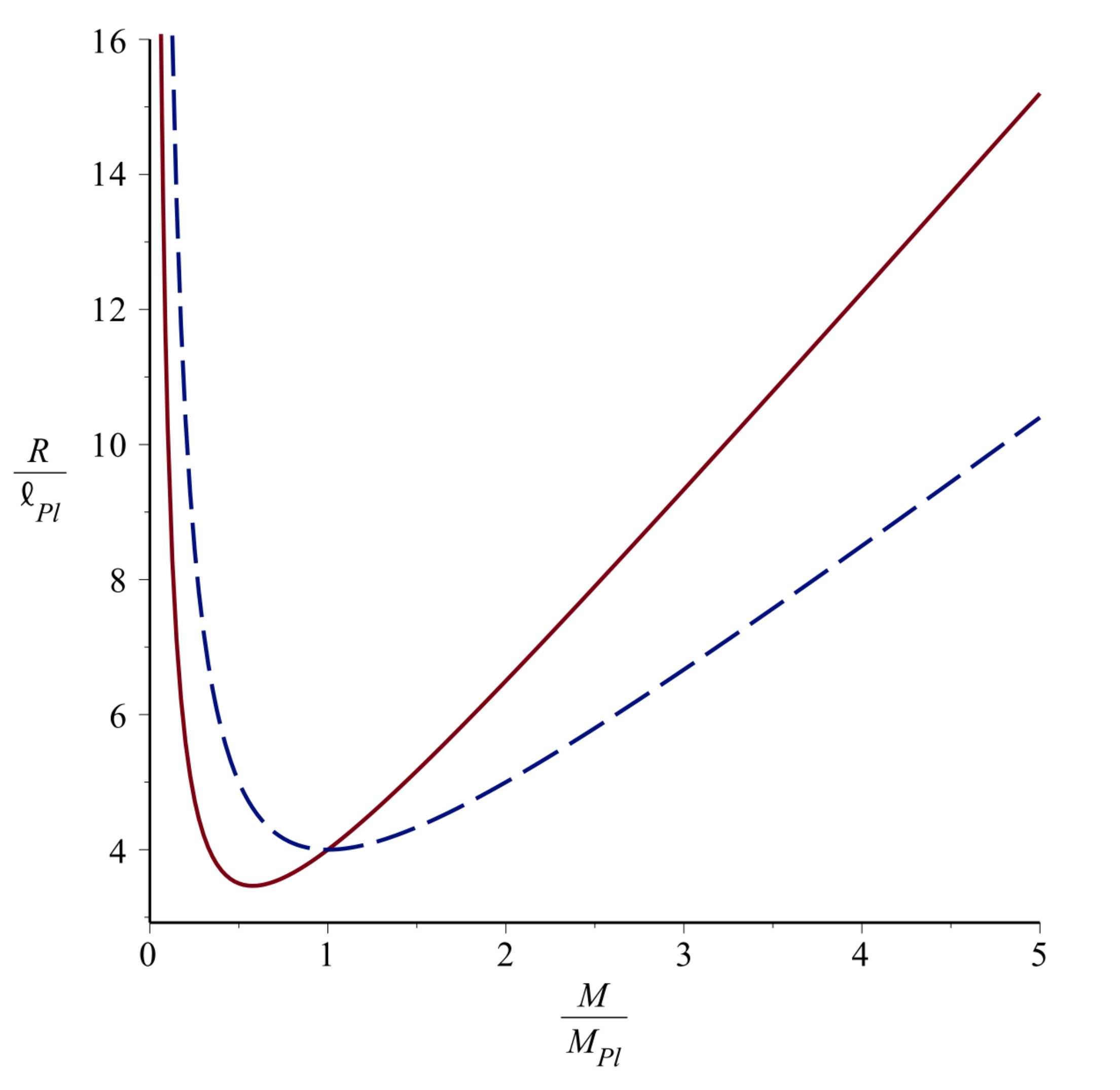}
\caption{Generalized Compton scale (red solid curve, Eq.~\ref{GUP2})
and Generalized Event Horizon scale (blue dashed curve, Eq.~\ref{GUP4B})
for the two ranges of $\alpha$ and $\beta$.  Left: $\alpha = 0.5, \beta = 0.7$; Right: $\alpha = 3, \beta = 2$. The two curves cross at $M_{\rm f}$ and $L_{\rm f}$ given by \eqref{scaleinter} and \eqref{scaleinterR}.}
\label{scalefig}
\end{figure}
\end{center}

This approach suggests that (in some sense) elementary particles {\it are} sub-Planckian black holes, \textit{i.e.} quantum mechanical objects with event horizons. 
However, while Fig.~\ref{modesto1} assumes that the curves $R_C'(M)$ and $R_S'(M)$ are the same, the gravitational correction to the Compton wavelength could in principle  be independent of the quantum mechanical correction to the Schwarzschild radius.  
Accordingly, we now explore an alternative scenario in which the generalized Compton wavelength and Schwarzschild radius are still defined by \eqref{GUP2} and \eqref{GUP4B}, respectively,  
but with $\alpha$ and $\beta$ being unrelated parameters. This is illustrated in Fig.~\ref{scalefig}. The novelty is that $\RS'$ is defined for $M<\Mpl$, which extends the duality relation $M\leftrightarrow M^{-1}$ exhibited by the GUP to the GEH. 
Thus $\RC'$ describes the compression phase of a particle that eventually collapses into a black hole, while $\RS'$ describes a black hole decaying  into a black particle (\textit{i.e.}  the opposite process). 
 
The expressions for $\RC'$ and $\RS'$ coincide at
\begin{equation}
M_\mathrm{f}=\sqrt{\frac{1-\beta}{2-\alpha}}\ \Mpl \, ,
\label{scaleinter}
\end{equation}
so this is the fundamental mass scale at which the transition between quantum mechanics and classical gravity takes place. 
Eq.~\eqref{scaleinter} has a solution if the radicand is positive, which corresponds to $\beta <1$ and $ \alpha < 2$ or $\beta > 1$ and $ \alpha >2$. 
We can also associate a fundamental length scale with the mass  $M_\mathrm{f}$:
\begin{eqnarray}
L_\mathrm{f}
=\left(\sqrt{\frac{2-\alpha}{1-\beta}}+\alpha \sqrt{\frac{1-\beta}{2-\alpha}}\right)  \ellp \nonumber\\
= \left( \frac{| 2-\alpha \beta| } {\sqrt{ (1-\beta)(2-\alpha)}} \right)  \ellp  \, ,
\label{scaleinterR}
\end{eqnarray}
where the last expression applies for the values of $\alpha$ and $\beta$ for which $M_{\rm f}$ and $L_{\rm f}$ are defined. 
If $\beta=1$ and $\alpha=2$, the condition $\RC'=\RS'$ becomes an identity.
This  corresponds to full reversibility of the aforementioned processes of particle compression and black hole decay. The BHUP correspondence itself requires $\alpha = 2 \beta$, since this ensures that $\Delta x$ can map into both $R_C'$ given by \eqref{GUP2} and $R_S'$ given by \eqref{GUP4B}.  

For $0<\beta<1$ and $0<\alpha<2$, as illustrated in Fig.~\ref{scalefig} (left),
we have a scenario in which the generalized Compton wavelength is bigger than the GEH in the sub-Planckian regime ($\RC'>
\RS'$) but  smaller than it in the trans-Planckian regime ($\RC'<
\RS'$). This means that particles have gravitational radii smaller than their actual size. Furthermore, the compression phase is halted by the formation of a black hole with horizon radius $2GM/c^2$ in the super-Planckian regime. However, the reverse process of black hole decay does not stop at the fundamental scale but leads to the formation of a sub-Planckian black hole, interpreted as a quantum mechanical object which is more compressed than an elementary particles of the same mass. Such black holes would correspond to the `quantum domain' in Fig. \ref{GUP1}. 

For $\beta>1$ and $\alpha>2$, as illustrated in Fig.~\ref{scalefig} (right), we have the opposite scenario, with the generalized Compton wavelength being smaller than the GEH in the sub-Planckian regime ($\RC'<
\RS'$) but bigger than it in the super-Planckian regime ($\RC'>
\RS'$). This means that the gravitational short-scale cut-off exceeds $2GM/c^2$ in the super-Planckian regime and that sub-Planckian black holes exhibit a \textit{neo}-semiclasical character, corresponding to the `classical domain' in the left side of the phase diagram.  

\subsection{Possible Link with 2D Gravity}

The scale $M_\mathrm{f}$ plays a crucial role
 in the above discussion. It arises even in the standard picture as the mass at which quantum gravity ``capsizes'' the power dependence of $M$ on length scale.
 We here offer a possible (if incomplete) explanation of how such a quantum gravity mechanism could work. First, 
we recall that (1+1)-D dilaton black holes naturally encode a $1/M$ term in their gravitational radii (see for instance \cite{robb1,robb2,robb3,robb4,robb5,robb6,robb7,jack1,jack2,jack3,grum1,grum2,jmpnprd}). We also recall that according to t'Hooft \cite{thooft}, gravity at the Planck scale might experience a (1+1)-D phase due to 
spontaneous dimensional reduction. Such a conjecture is further supported by studies of the fractal properties of a quantum spacetime. Because of its intrinsic graininess and the local loss of resolution, quantum spacetime is expected to behave like a fractal at the Planck scale. Accordingly, one can consider the spectral dimension, an effective indicator of the fractal dimension as perceived in a random walk or diffusive process. In several quantum gravity formulations, the spacetime dimension at the Planck scale decreases and approaches two \cite{cdt1,cdt2,cdt3,modestocqg,dario,lmpn,pnes,Carlip2009,calcagni,ds,jrmds1,jrmds2,
rinaldi2,jrmbh}.

As a result, the decaying black hole might experience a transition to a temporary (1+1)-D phase when approaching the Planck scale. At this point  the Komar energy can be defined in much the same way as for dilaton black holes \cite{jmpnprd, jmpnepj}:
\begin{equation}
M\sim \int dx\sqrt{g^{(1)}} \, n_i ^{(2)}T^{\ i}_{ 0} \, ,
\label{qbhkomar2d}
\end{equation}
where $g^{(1)}$ is the determinant of the spatial section of $g^{ij}$, the effective 2D quantum spacetime metric, and $^{(2)}T^{\ i}_{ 0}$ is the dimensionally reduced energy-momentum tensor. 
As noted in \cite{jmpnepj}, once black holes undergo the transition to (1+1)-D geometry, they are no longer sensitive to any Planck scale discrimination between classical and quantum gravity. In 2D the coupling constant is dimensionless and black holes can naturally be extended to the sub-Planckian sector.

\section{A new quantum-corrected black hole solution}

On the basis of the discussion of the previous section, we posit a quantum correction to the Schwarzschild metric of the form
\bea
ds^2 & = &F(r) dt^2 - F(r)^{-1} dr^2 - r^2 d\Omega^2 \nonumber \\
F(r) &=& 1-\frac{2}{\Mpl^2} \frac{M}{r}\left(1+\frac{\beta}{2} \frac{\Mpl^2}{M^2}\right)  \, ,
\label{newmetric}
\eea
where we henceforth use units with $\hbar = c =1$. 
The metric \eqref{newmetric} is Schwarzschild-like, in the sense that
the modification factor $1+\beta \Mpl^2/(2M^2)$ is coordinate-independent.  It is a vacuum solution with the usual tensor and scalar behaviour.  Although we have not removed the singularity, we will show that it can never be reached. 
We note a possible connection with the energy-dependent metrics proposed in the framework of ``gravity's rainbow''\cite{rainbow}.
We  also note that such a way of tackling the problem resembles the findings of Camacho, who distinguished the bare and renormalized mass in QFT in presence of stochastic metric fluctuations \cite{camacho}.

\begin{center}
\begin{figure}[h]
\includegraphics[scale=0.4]{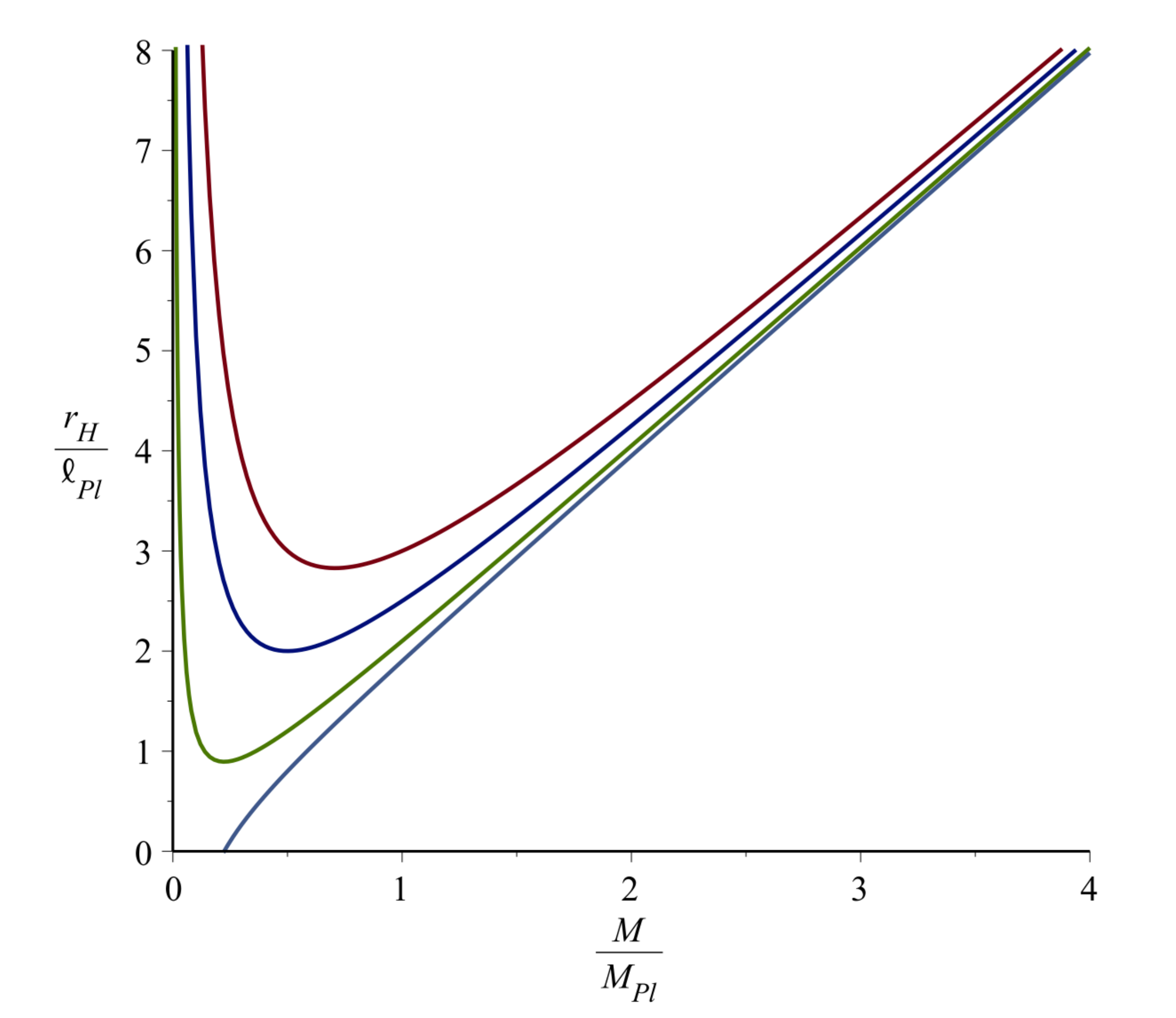}
\caption{Horizon radius
 (\ref{horizon}) 
 as a function of mass $M/\Mpl$ for $\beta = 1$ (red, top), $\beta = 0.5$ (blue, second down), $\beta = 0.1$ (green, third down) and $\beta = -1$ (blue, bottom).  In the last case, the horizon vanishes when $M = \sqrt{|\beta|/2}\Mpl$ and is defined only for $M > \sqrt{|\beta|/2}\Mpl$.}
\label{fig1}
\end{figure}
\end{center}

The horizon size for the metric \eqref{newmetric} is given by
\beq
\rH = \frac{2}{\Mpl^2} \left( \frac{M^2+\frac{\beta}{2}\Mpl^2}{M}\right) = \RS'~~,
\label{horizon}
\eeq
as anticipated.
The relationship between $r_\mathrm{H}$ and $M$ is shown in Fig.~\ref{fig1}.
In the three mass regimes, we find 
\bea
M \gg \Mpl & \Longrightarrow & \rH \approx \frac{2M}{\Mpl^2} \label{bigr}\\
M \approx \Mpl & \Longrightarrow & \rH \approx \frac{2+\beta}{\Mpl} \label{ellpr}\\
M \ll \Mpl & \Longrightarrow & \rH \approx \frac{\beta}{M} \label{smallr}~~~.
\eea
The first expression is 
the standard Schwarzschild radius  The intermediate expression gives a minimum 
of order $\ellp$, so the Planck scale is never actually reached for $\beta>0$, in agreement with the principle of inaccessibility of the singularity.  
The last expression scales inversely with mass and resembles the Compton wavelength.

The metric exhibits dimensional reduction: in the classical regime, it is the $(3+1)$-D Schwarzschild solution, while in the sub-Planckian limit it resembles the $(1+1)$-D limit of general relativity \cite{robb1}, in which the horizon radius also scales  as $M^{-1}$.
Note that the actual dimension of the manifold is still four; the form of the metric circumvents the necessity of introducing a specific dynamical dimensional reduction mechanism. In other words, the $M^{-1}$ dependence is just the memory of a temporary $(1+1)$-D phase at the Planck scale.

For $\beta>0$. Eq.~(\ref{horizon}) can be inverted to give the two masses associated with a given horizon radius $\rH$.
It is instructive to write these as 
\beq
\frac{2M_\pm}{\Mpl^2} = \frac{\rH \pm \sqrt{\rH^2 - 8 \beta \ellp^2}}{2} \, ,
\label{mpm}
\eeq
which gives a minimum radius
\beq
r_{\rm min} = 2\sqrt{2 \beta} \, \ellp, \quad
\eeq
and an associated mass
\beq
M(r_{\rm min}) = \sqrt{\beta/2} \, \Mpl \, .
\eeq
This is illustrated in Fig.~\ref{fig1}.   Since
$\rH \geq r_{\rm min}$, the singularity can never be externally probed, which is in agreement with the claimed ultraviolet self-complete character of gravity \cite{dvali, piero4}.

For $\beta < 0$, only the $M_+$ solution applies and from (\ref{horizon}) this admits an horizon
 only for $M>\sqrt{|\beta|/2} \, \Mpl$.
This suggests there are sub-Planckian black holes in this case only for $0 > \beta> -2$. It has been claimed that the form of the GUP predicted by loop quantum gravity \cite{ashtekar_1, ashtekar_2} implies $\beta <0$ because the lattice structure of spacetime reduces the fluctuations. However, the LBH solution can still have $\beta > 0$ in the quadratic GUP case \cite{cmp}.  
It should be noted that solutions with $\beta < 0$ are not endowed with a ultraviolet cutoff. In the limit $\rH\to0$ one finds $M\to\sqrt{|\beta|/2} \, \Mpl$, but there is still a divergent temperature (see below). Therefore the situation is not different from 
the standard evaporation scenario, although it could have interesting thermodynamical aspects.

We can also express the metric in terms of the Newtonian potential
\beq
F(r) = 1+2\Phi(r), \quad
\Phi = -\frac{1}{\Mpl^2} \frac{M}{r} \left(1+\frac{\beta \Mpl^2}{M^2}\right) \, .
\eeq
We then find 
\beq
\Phi(r) \sim -\frac{\beta}{rM}
\eeq
in the limit $M \ll \Mpl$. This still corresponds to an inverse-square force law but it is large at small masses.  In the limit $M \gg \Mpl$, we recover the classical Newtonian potential.
This is reminiscent of the ``strong'' gravity model of the 1970s, which was introduced in an attempt to model elementary particles as black holes \cite{holzhey1,holzhey2,holzhey3,holzhey4}. With ordinary gravity, an elementary particle is much larger than its Schwarzschild radius, the factor being  $(\Mpl/m_{\mathrm p})^2 \sim 10^{40}$ for a proton. However, it could be interpreted as a black hole if the gravitational constant were increased by this factor. So this is another interpretation of the BHUP correspondence.  

\section{GUP and Black Hole Thermodynamics}
Given the metric (\ref{newmetric}), we now explore the thermodynamics of the black hole solutions in the three limits considered above (super-Planckian, trans-Planckian and sub-Planckian).
Since the exact nature of the physics is unknown in the trans-Planckian and sub-Planckian regimes, we use two approaches. We first 
calculate the temperature using the traditional Adler approach \cite{Adler_1,Adler_2,Adler_3,Adler_4} and then calculate it using the surface gravity argument. These give slightly different results, although they agree asymptotically. 

\subsection{Adler Method}
Let us first recall the link between black hole radiation and the HUP~\cite{hawking1,hawking2}.  This arises because we can obtain the black hole temperature for $M \gg \Mpl$ by 
identifying $\Delta x$ with the Schwarzschild radius and $\Delta p$ with a multiple of the black hole temperature:
\begin{eqnarray}
kT = \eta c \Delta p = \frac{ \eta \hbar  c}{\Delta x} = \frac{\eta \hbar c^3}{ 2 G M} \, .
\label{temper}
\end{eqnarray}
This gives the precise Hawking temperature 
if we take $\eta = 1/(4\pi)$. 
The second equality in (\ref{temper}) relates to the emitted particle and assumes that $\Delta x$ and $\Delta p$ satisfy the HUP (one can include the factor of $2$ in the HUP by rescaling $\eta$.).  The third equality relates to the black hole and assumes that $\Delta x$ is the Schwarzschild radius.  Both these assumptions require $M \gg \Mpl$, so we now generalize this argument.

\begin{figure}[b]
\includegraphics[scale=.42]{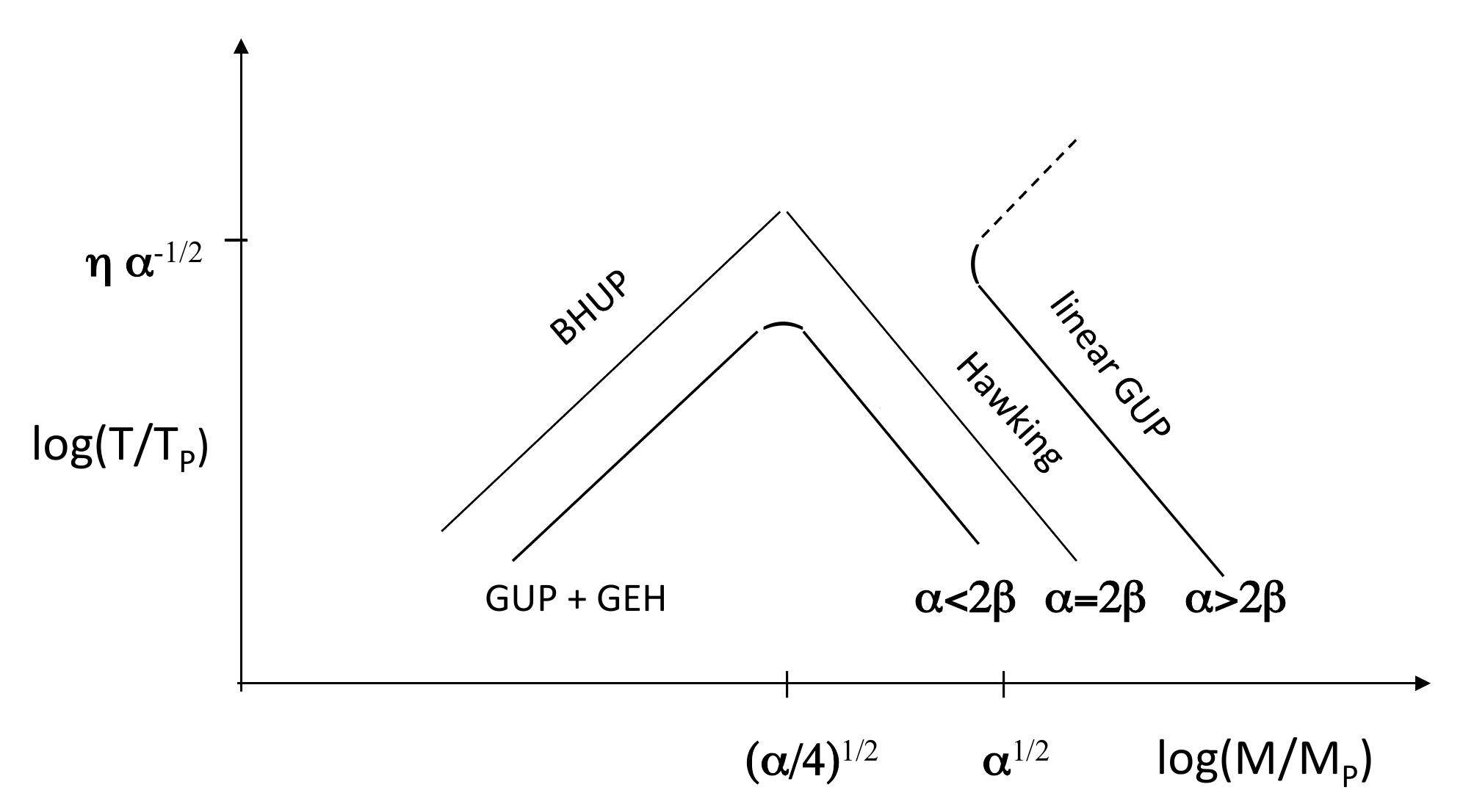}
\caption{Comparing black hole temperature (\ref{rainbowtem}) to that predicted by Hawking, linear GUP and BHUP correspondence. }
\label{temp}     
\end{figure}

Adler {\it et al.} \cite{Adler_1, Adler_2, Adler_3, Adler_4} calculate the modification required
if  $\Delta x$ is still 
associated  with the Schwarzschild radius but $\Delta p$ and $\Delta x$ are related by the linear form of the GUP. In this case, 
identifying the size of the black hole with the wavelength of the emitted radiation gives
\begin{equation}
\frac{2GM}{c^2} = \frac{\hbar \eta c}{kT} + \frac {\alpha G k T}{\eta c^4}
\end{equation}
where $\alpha$ is the parameter in (\ref{GUP1}). This 
leads to a temperature 
\begin{equation}
T = {\eta M c^2 \over \alpha k} \left(1- \sqrt{1- \frac{\alpha \Mpl^2}{M^2}} \right) \, ,
\label{adlertemp2}
\end{equation}
giving a small perturbation to the standard Hawking temperature,
\begin{equation}
T \approx
{\eta \hbar c^3 \over 2 G k M}  \left[ 1 + {\alpha \Mpl^2 \over 4M^2} \right] \, ,
\end{equation}
for $M \gg \Mpl$. However, the exact expression 
becomes complex when $M$ falls below $\sqrt{\alpha} \, \Mpl$, corresponding to a minimum mass and a maximum temperature.  This is indicated by the curve on the right in Fig.~\ref{temp}.

In the present model,  \eqref{newmetric} suggests that the temperature is still given  by (\ref{adlertemp2}) providing $M$ is interpreted as the ADM mass \eqref{newmass}. However, since this has a minimum value of $\sqrt{\beta/2} \, \Mpl$, 
the temperature reaches a maximum and then decreases  for $\alpha < 2 \beta$ rather than going complex. We can also express $T$ in terms of $M$ 
by identifying the size of the black hole with the wavelength of the emitted radiation:
\begin{eqnarray}
 \left( \frac {\hbar \eta c}{kT} \right) + \left( \frac{\alpha \ellp^2 kT}{\hbar \eta c} \right)
= \left( \frac{\hbar \beta}{Mc} \right) + \left( \frac{2GM}{c^2}\right) \, .
\end{eqnarray}
This implies
\begin{eqnarray}
kT = \frac{\eta M c^2}{\alpha} f(M, \alpha, \beta)
\label{rainbowtem}
\end{eqnarray}
where the function $f$ is real for $\alpha < 2 \beta$ and given by 
\begin{eqnarray}
f(M, \alpha, \beta) = 
 1 + {\beta \over 2} \left({\Mpl \over M}\right)^{2} \\ \nonumber
\pm  \sqrt{1+ (\beta - \alpha)\left( {\Mpl \over M} \right)^{2} + {\beta^2 \over 4} \left({\Mpl \over M}\right)^{4} } \, .
\end{eqnarray}
This can be approximated by
\begin{equation}
T \approx {\eta \hbar c^3 \over 2 G k M} \left[ 1 - \left( \frac {2\beta - \alpha}{4}\right)  \left( {\Mpl \over M} \right)^{2} \right]  
\label{super}
\end{equation}
for $M \gg \Mpl$ and
\begin{equation}
T \approx {\eta M c^2 \over k \beta} \left[ 1 - \left( \frac{2\beta - \alpha}{ \beta^2}\right)  \left( {M \over \Mpl} \right)^{2} \right]   
\label{sub}
\end{equation}
for $M \ll \Mpl$. The overall behaviour of $T$ is shown by the lower curve in Fig.~\ref{temp}.

In the special case $\alpha = 2 \beta$, associated with the BHUP correspondence, one obtains the solution \cite{carr_2} 
\begin{eqnarray}
kT = \mathrm{min} \left[ \frac{\hbar \eta c^3}{2GM} \,  , \, \frac{2\eta Mc^2}{\alpha} \right]  \, .
\label{mess3}
\end{eqnarray}
This is indicated by the middle curve in Fig.~\ref{temp}. The first expression in \eqref{mess3} is the {\it exact} Hawking temperature, 
with no small correction term, but
one must cross over to the second expression below $M = \sqrt{\alpha /4} \, \Mpl$ in order to avoid the temperature going above the Planck value $\Tpl = \Mpl c^2/k$. The second expression in \eqref{mess3}
can be obtained by putting  $\Delta x = \hbar \beta / (Mc)$ in (\ref{temper}).
Since $T < \Tpl$ for all $M$, the second equality in (\ref{temper}) still applies to a good approximation.

Equation~(\ref{mess3}) also applies for loop black holes \cite{cmp} but the physical interpretation is somewhat different in that case.  This is because
there are two different asymptotic spaces in the LBH solution and the temperature of a sub-Planckian black hole goes as $M^3$ in our space and $M$ in the other space. (The heuristic argument only gives the temperature in the space which contains the event horizon.)
In the present case, 
there is only one asymptotic space and the different $M$-dependences for $M < \Mpl$ and  $M > \Mpl$ just arise from the metric's limiting behaviour.

\subsection{Surface Gravity Method}

\begin{center}
\begin{figure}[h]
\includegraphics[scale=0.4]{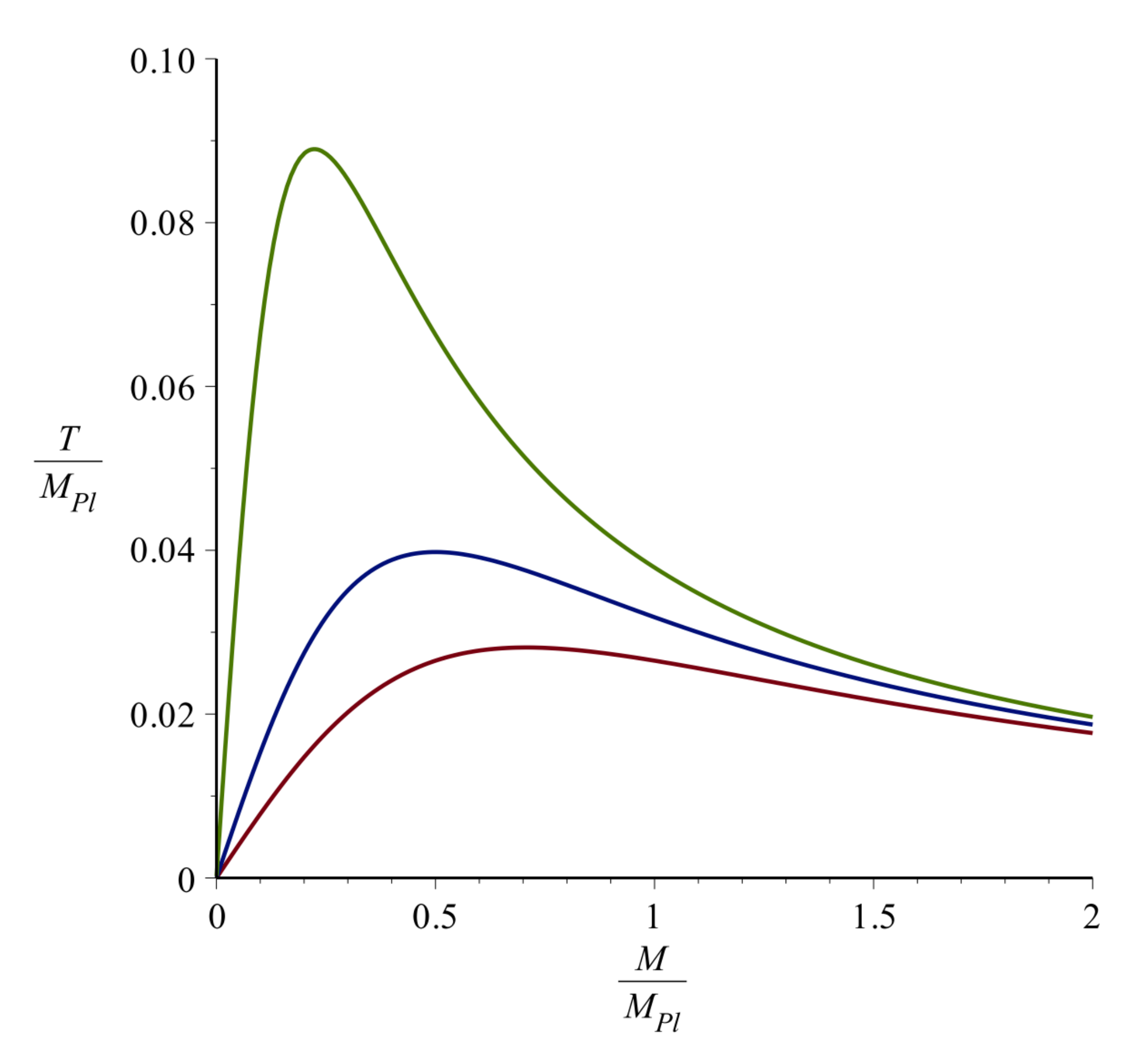}
\caption{Hawking temperature (\ref{hawktemp}) implied by surface gravity argument as a function of mass for $\beta = 1$ (red, bottom), $\beta = 0.5$ (blue, middle), and $\beta = 0.1$ (green, top).  The temperature reaches a maximum at (\ref{mplm}), after which the heat capacity becomes
positive and the black hole cools to a zero-temperature configuration, at which point it evaporates completely.}
\label{fig2}
\end{figure}
\end{center}

One can also use another argument to obtain the black hole temperature. If this is determined by the black hole's surface gravity \cite{hawking1,hawking2}, one expects
\beq
T = \frac{\kappa}{2\pi} = \left. \frac{1}{4 \pi} \frac{d g_{tt}}{dr} \right|_{r=\rH} \, ,
\eeq
which gives 
\beq
T = \frac{\Mpl^2}{8\pi M(1+\beta \Mpl^2/2M^2)} \, .
\label{hawktemp}
\eeq
This temperature is plotted in Fig.~\ref{fig2} and the limiting behaviour in the 
different mass regimes is as follows: 
\bea
M \gg \Mpl & \Longrightarrow & T \approx \frac{\Mpl^2}{8\pi M} \left[ 1- \beta  \left( {\Mpl \over M} \right)^{2} \right]   \label{bigm}\\
M \approx \Mpl & \Longrightarrow & T \approx \frac{\Mpl}{8\pi(1+\beta/2)} \label{mplm}\\
M \ll \Mpl & \Longrightarrow & T \approx \frac{M}{4\pi \beta} \left[ 1- \frac{1}{\beta}  \left( {M \over \Mpl} \right)^{2} \right] \, . \label{smallm}
\eea
The large $M$ limit (\ref{bigm}) is the usual Schwarzschild temperature and is unstable.  However, as the black hole evaporates, it reaches a maximum temperature (\ref{mplm}).  Below this point, the thermodynamics stabilizes and the temperature approaches zero as $M \rightarrow 0$. The behaviour for $M \ll M_{Pl}$ again replicates the temperature of a $(1+1)$-D black hole, since this also scales as $M$ \cite{jmpnepj}.
As discussed in Sec.~V, this leads to {\it effectively} stable relics which might in principle provide the dark matter~\cite{poly_1, poly_2, poly_3, poly_4,  poly_6, poly_7,jrmbh}.  

If one were to plot (\ref{hawktemp}) in Fig.~\ref{temp}, it would be similar to the lower curve but not identical. 
The temperatures given by (\ref{rainbowtem}) and (\ref{hawktemp}) agree to 1st order but not to 2nd order.
This is because comparison with (\ref{super}) and (\ref{sub}) shows that the higher order terms agree only for $\alpha = - 2\beta$ in the $M \gg M_P$ regime and $\alpha = \beta$ in the $M \ll M_P$ regime. Therefore they cannot agree in both asymptotic limits. 
The same problem arises in the LBH solution and can only be resolved by arguing that the definition of many physical quantities must be modified as one approaches the Planck limit \cite{cmp}. 
The qualitative implications are similar in both cases.

\begin{center}
\begin{figure}[h]
\includegraphics[scale=0.4]{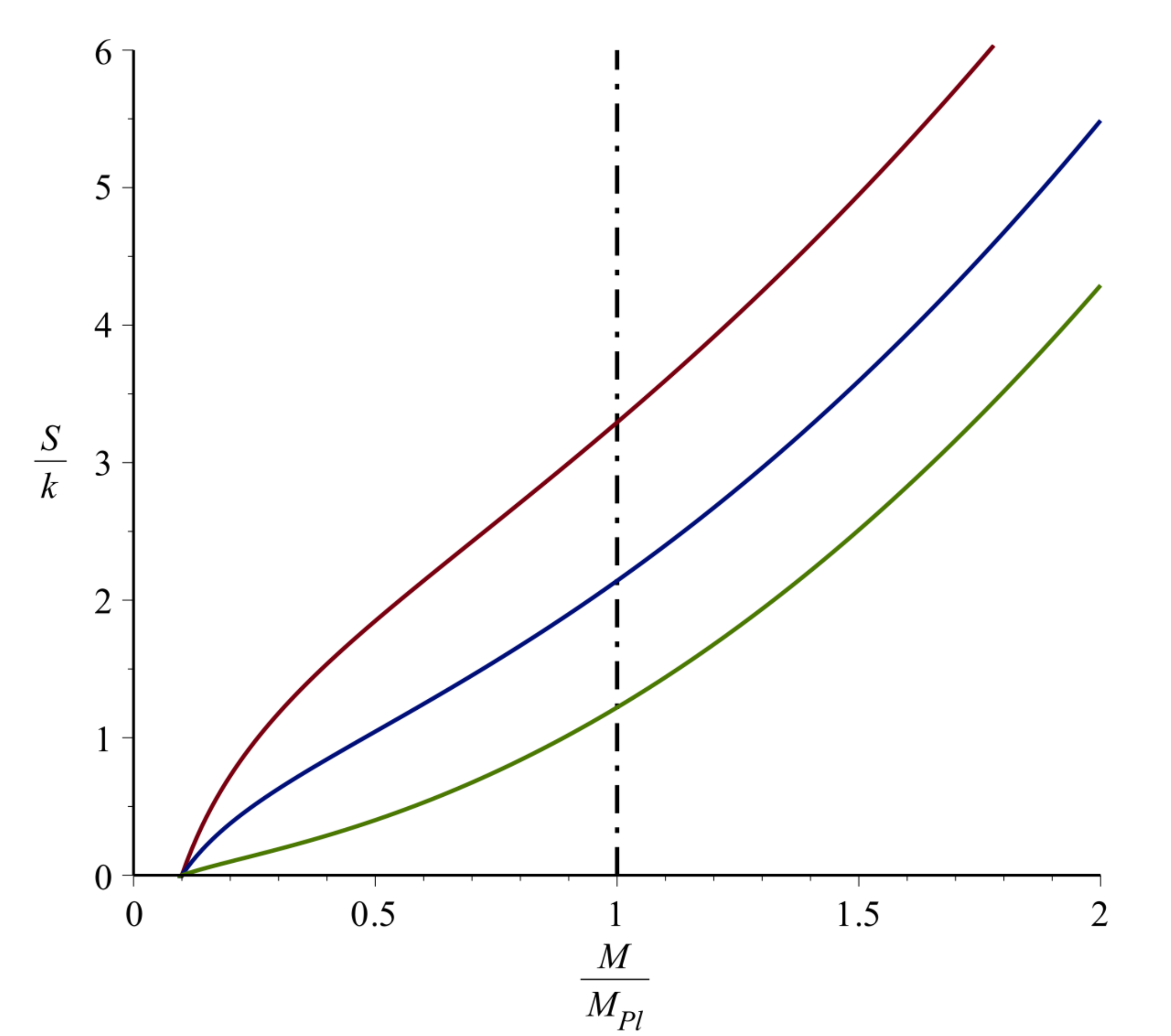}
\caption{Black hole entropy (\ref{entropy}) as a function of mass for $\beta = 1$ (red, top), $\beta = 0.5$ (blue. middle), and $\beta = 0.1$ (green. bottom).  The Planck scale is marked by the vertical hatched line ($M=\Mpl$).  For $M>\Mpl$, the entropy scales classically as $S\sim M^2$ (plus sub-leading terms), while for $M<\Mpl$ it scales as $S\sim \log(M/M_0)$ and vanishes at $M=M_{0}$ (set as $ 0.1$ in the plot). }
\label{fig3}
\end{figure}
\end{center}

If the temperature is given by (\ref{hawktemp}), the black hole entropy can be calculated in the usual fashion as
\beq
S = \int_{M_0}^M\frac{dM^\prime}{T(M')} = 4\pi k \left(\frac{M^2}{\Mpl^2}-\frac{M^2_{0}}{\Mpl^2} + \beta \ln \frac{M}{M_{\rm 0}}\right)
\label{entropy}
\eeq
where $M_0 < \Mpl$ is some lower bound of integration. This expression is plotted in Figure~\ref{fig3}.
If the temperature is given by (\ref{rainbowtem}), the expression for $S$ is more complicated but still has the same asymptotic limits. In both cases, the presence of a logarithmic correction is consistent with the entropy of a $(1+1)$-D Schwarzschild spacetime \cite{robb1}.  It also agrees with the notion that $(1+1)$-D black holes are naturally quantum objects \cite{jmpnepj}, emerging here via the dependence of $\rH$ on $M$. 

This is in agreement with a model-independent feature emerging from a variety of approaches to quantum gravity including string theory \cite{vafa}, loop quantum gravity \cite{carlo}, ultraviolet gravity self-completeness \cite{piero4} and other formulations based on mechanisms for counting microstates \cite{robb98,kaul}.    The lower bound of integration represents some minimal configuration of the black hole, which is possibly related to a quantity to be derived in the following section.
The classical limit of (\ref{entropy}) has the expected form $S \propto M^2$.
The approach of a mass redefinition \eqref{newmass}
is inspired by the GUP's duality, and the associated black hole thermodynamics obviously encodes this.  The above asymptotic agreements are encouraging, since they suggest that the derived thermodynamical limits are largely model-independent and so can be assumed with a fair degree of confidence.

\section{Observational Consequences}
\label{obscons}

If the black hole temperature is given by (\ref{hawktemp}), its luminosity ($L \propto \rH^2T^4$)
can be written as  
\beq
L 
= \gamma^{-1} M^{-2} \left( 1+\frac{ \beta\Mpl^2}{2M^2} \right)^{-2} 
\eeq
where $\gamma \sim t_\mathrm{Pl}/\Mpl^3$. If the primordial black hole (PBH) forms at time $t_\mathrm{i}$ with mass $M_\mathrm{i}$, its mass then evolves according to 
\bea
 t - t_\mathrm{i}&
= &\frac{\gamma}{3} (M_\mathrm{i}^3-M^3) +  \gamma \beta (M_\mathrm{i} - M)\Mpl^2\nonumber\\ &-& \frac{\gamma \beta^2}{4} \left( \frac{1}{M_\mathrm{i}} - \frac {1}{M} \right)\Mpl^4 \, .
\label{evap}
\eea
Although the black hole loses mass on a timescale 
\beq
\tau  \sim M/L \sim \gamma M^{3}(1+ \beta \Mpl^2/2M^2)^{2} \, ,
\eeq
it never evaporates entirely because the last term in  (\ref{evap}) implies that it takes an infinite time for $M$ to go to zero. Nevertheless, there are two values of $M$ for which $\tau$ is comparable to the age of the Universe ($t_0 \sim 10^{17}$s) and these turn out to have physical significance. One is super-Planckian,
\beq
M_* \sim (t_0/ \gamma)^{1/3} \sim (t_0/\tpl)^{1/3} \Mpl \sim 10^{20} \Mpl \sim 10^{15}~\rm g \, ,
\eeq
and the other sub-Planckian, 
\beq
M_{**} \sim  \beta^2 (\tpl/t_0) \Mpl \sim 10^{-60} \Mpl \sim 10^{-65}~\rm g \, ,
\eeq
where we have assumed 
 $\beta \sim {\cal O}(1)$ in the last two estimates.
The mass $M_*$ is the standard expression for the mass of a primordial black hole (PBH) evaporating at the present epoch. This still applies except that the mass does not shrink all the way to zero, 
\textit{i.e.}, for $M_\mathrm{i} \gg \Mpl$ the usual Hawking lifetime $\tau \propto M^3$ must be interpreted as the time to shrink to the mass $\Mpl$. Thereafter the mass quickly evolves to the value $M_{**}$ at which $L t_0 \sim M$.
Although this mass-scale may seem implausibly small, it arises naturally in some estimates for the photon or graviton mass ({\it e.g.} 
in the work of Mureika and Mann~ \cite{rbmjm}). It effectively specifies the lower integration bound in (\ref{entropy}), {\it i.e.} $M_{**} = M_0$.

\if
This gives the limiting behaviours:
\bea
M \gg \Mpl & \Longrightarrow & \tau \sim \tpl\left(  \frac{M}{\Mpl} \right)^3   \label{largem}\\
M \sim \Mpl & \Longrightarrow & \tau \sim \tpl \label{medm}\\
M \ll \Mpl & \Longrightarrow & \tau \sim \tpl\left( \frac{M}{\Mpl} \right)^{-1} \label{littlem} \, .
\eea
\fi

For $M_\mathrm{i} \gg \Mpl$, the mass at the present epoch ($t_0$) has reduced to 
\beq
M = [M_\mathrm{i}^3  - M_*^3 (1-t_\mathrm{i}/t_0)]^{1/3} \, .
\eeq
Hence $M \approx M_\mathrm{i}$ for $M \gg M_*$ (\textit{i.e.} it is unchanged) and $M$ falls into the sub-Planckian regime for $M_i < M_*$. For $M_\mathrm{i} \ll \Mpl$, the mass at the present epoch has reduced to 
\beq
M = \frac{M_\mathrm{i}}{1+M_\mathrm{i}(t_0-t_\mathrm{i})/(\beta^2 \Mpl t_{\rm Pl})} \approx\frac{M_\mathrm{i}}{1+M_\mathrm{i}/M_{**}}  \, .
\eeq
Hence $M \approx M_\mathrm{i}$ for $M \ll M_{**}$ (\textit{i.e.} it is unchanged) and $M \approx M_{**}$ for $M_* \gg M_i \gg M_{**}$.
It is unclear whether black holes can form with sub-Planckian mass; probably they can only evolve there by evaporation from the super-Planckian regime. 

The mass cannot actually reach the value $M_{**}$ at the present epoch because of the effect of the cosmic microwave background (CMB). This is because the black hole temperature is less than the CMB temperature ($T_{\rm CMB}$), suppressing evaporation altogether, below an epoch-dependent mass
\beq
M_{\rm CMB} = 10^{-36}(T_{\rm CMB}/3K) \,  \mathrm{g} \, .
\eeq
Accretion is expected to be unimportant \cite{ch1974}, so the PBH mass effectively freezes at this value.  This leads to {\it effectively} stable relics which might in principle provide a candidate for dark matter~\cite{poly_7,jrmbh}. The relic PBH mass decreases with time but the current value is around $10^{-4}$eV, which is also the mass-scale associated with the dark energy. 

\begin{center}
\begin{figure}[h]
\includegraphics[scale=0.5]{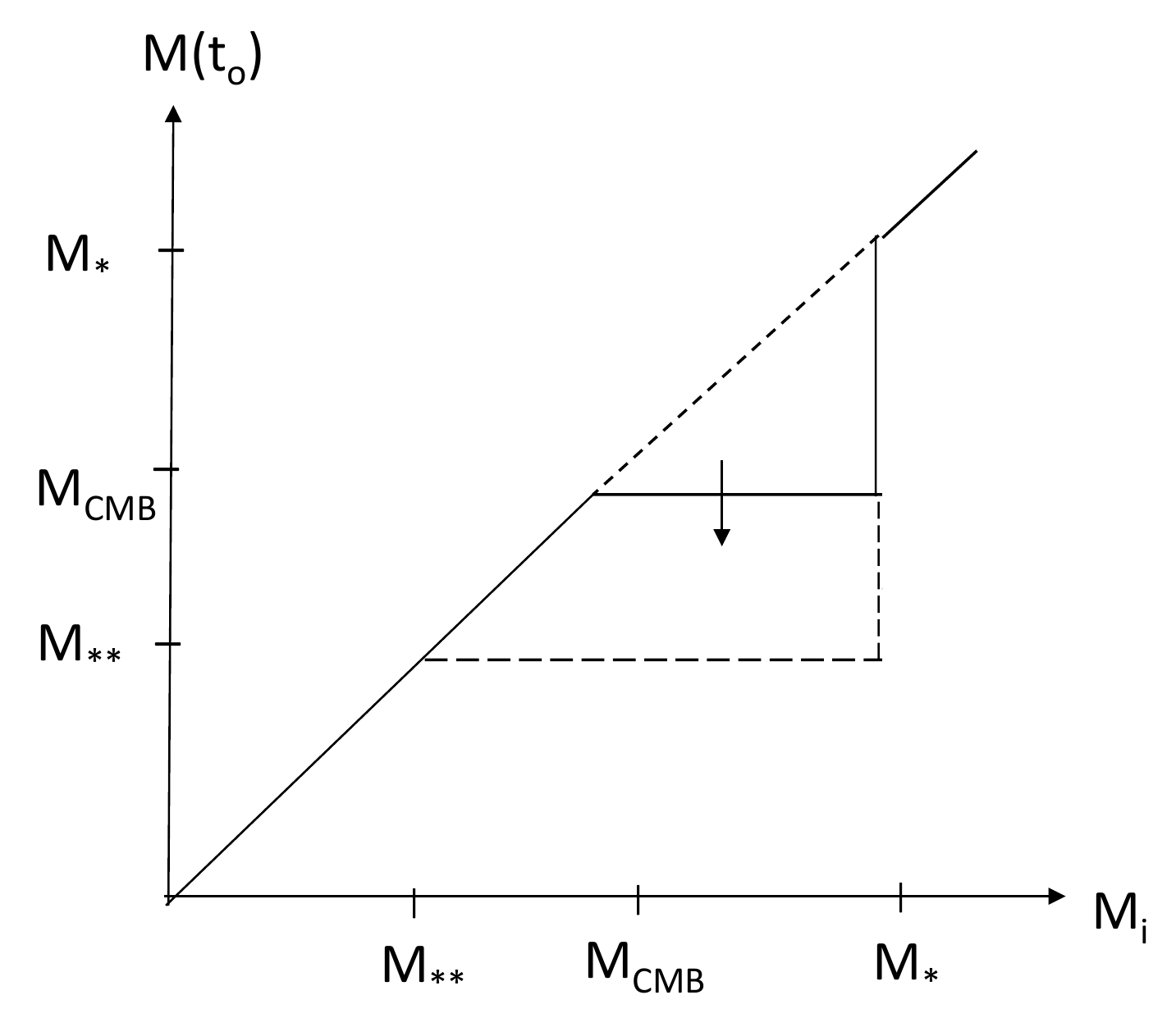}
\caption{Current black hole mass $m \equiv M(t_0)$ as a function of initial mass $M_i$, showing discontinuities at $M_*$ and $M_{**}$ (where evaporation timescale equals age of Universe) and $M_{\rm CMB}$ (where black hole has CMB temperature, this decreasing with epoch). }
\label{fig8}
\end{figure}
\end{center}

The value of the PBH mass at the present epoch can be approximated as $M_\mathrm{i}$ for $M_i<M_{\rm CMB}$ and $M_i>M_*$ but as $M_{\rm CMB}$ for $M_{\rm CMB}<M_i<M_*$,
as illustrated in Fig.~\ref{fig8}. The discontinuities at $M_*$ and $M_{\rm CMB}$ would obviously be smoothed out in a more precise treatment.
The mass $M_*$ 
corresponds to a black hole radius 
\beq
\rH \sim (t_0/\tpl)^{1/3} \ellp \sim r_{\rm p} \sim 10^{-13}\rm cm
\eeq
and temperature
\beq 
T \sim (t_0/\tpl)^{-1/3} \tpl \sim m_{\rm p} \sim 10^{12} \rm K \, , 
\eeq
where $m_{\rm p}$ and $r_{\rm p}$ are the mass and radius of the proton.  These quantities arise in the expressions for the size and mass because of the Dirac ``large number'' coincidence ($t_0 \sim \alpha_{\rm G}^{-3/2}\tpl$ where $\alpha_{\rm G} = Gm_{\rm p}^2/\hbar c \sim 10^{-38}$ is the gravitational fine structure constant). One expects such PBHs to generate a background of 100~MeV gamma-rays and cosmic rays, the constraints on such PBHs being well explored~\cite{carr3}. 

The mass $M_{**}$ corresponds to a radius
\beq
\rH \sim (t_0/\tpl) \ellp \sim 10^{60}\ellp  \sim 10^{27} \rm cm
\eeq
and temperature
\beq
T \sim (t_0/\tpl)^{-1} \tpl \sim 10^{-28}\rm K  \, .
 \eeq
The first expression corresponds to the current cosmological horizon size and the second to the Hawking temperature for a black hole with the mass of the Universe, so the physical significance of such an extreme sub-Planckian black hole
 (which we term a ``minimon'') 
is unclear. 
It might seem unlikely that such objects could have observational consequences but surprisingly strong constraints are associated with gravitational and Dvali-Gabadadze-Porrati (DPG) effects on the scale of clusters ~\cite{nieto}. 

\section{Discussion}

The usual argument for the BHUP correspondence starts with the GUP in the sub-Planckian regime, and then extrapolates to black holes in the super-Planckian regime to infer an expression for the GEH. We have presented a different approach, in which one starts with a new type of black hole solution in the super-Planckian regime and then extrapolates down to the sub-Planckian regime to derive the GUP.  In this sense, the GUP is encoded in  the metric itself and 
the duality in the momentum uncertainty $\Delta p \leftrightarrow 1/\Delta p$ suggests a mass duality $M \leftrightarrow 1/M$ in the gravitational sector. 

We have also discussed the possibility that the GUP and GEH lines in the ($M,R$) diagram are different. Even in this case, one still has the possibility of sub-Planckian black holes and this raises the issue of a possible relationship between  black holes and elementary particles.  There is also some connection here to an approach based on extending the de Broglie relations to the gravitational domain~\cite{lake}. 

We have presented a comprehensive analysis of the thermodynamics of GUP black holes in the super-Planckian to sub-Planckian regimes. Due to the lack of an exact quantum theory of gravity, we have approached the problem in two ways.  First, we have followed the standard method \cite{Adler_1, Adler_2, Adler_3, Adler_4} of assuming the GUP {\it a priori} and then deriving the implied thermodynamic characteristics.  Second, we have calculated the temperature from the surface gravity  of the new solution. The predicted temperatures have the same form for the limiting cases $M \gg \Mpl$ and $M \ll \Mpl$, indicating that these features may be model independent.  

One novel aspect of this interpretation of the GUP is that it implies effective dimensional reduction.  This is because  the relationship between the horizon radius and black hole mass in the super-Planckian and sub-Planckian regimes corresponds to $(3+1)$-D and $(1+1)$-D gravity, respectively. Indeed, the algebraic form of the black hole parameters in $(1+1)$-D are the same as its quantum mechanical characteristics.  In this sense, two-dimensional gravity is naturally quantum mechanical in that the two theories become effectively indistinguishable. Instead of requiring two separate theories to govern the classical and quantum regimes ({\it i.e.} general relativity and quantum mechanics), the problem is recast into a single theory, where gravitation governs both, but in different effective spacetime dimensions.

\begin{acknowledgments}
JM and PN would like to thank Queen Mary University of London, where this work was done, for its generous hospitality.   The work of JM was supported by a Frank R. Seaver Research Fellowship from Loyola Marymount University.  The work of PN was supported by the
German Research Foundation (DFG) grant NI 1282/2-1, and partially by the Helmholtz International Center for FAIR within the framework of the LOEWE program (Landesoffensive zur Entwicklung Wissenschaftlich-
\"{O}konomischer Exzellenz) launched by the State of Hesse and partially by the European COST action MP0905 ``Black Holes in a Violent Universe''.

\end{acknowledgments}


\begin{thebibliography}{99}

\bibitem{Adler_1} R. J. Adler and D. I. Santiago, 
Mod. Phys. Lett. {\bf A14}, 1371 (1999).
\bibitem{Adler_2}R. J. Adler, P. Chen and D. I. Santiago, 
Gen. Rel. Grav. {\bf 33}, 2101 (2001).
\bibitem{Adler_3} P. Chen and R. J. Adler, 
Nucl.Phys.Proc.Suppl. \textbf{124}  103 (2003). 
\bibitem{Adler_4}R. J. Adler, 
Am.\ J.\ Phys.\  {\bf 78}, 925 (2010).


\bibitem{ashtekar_1}
A. Ashtekar, S. Fiarhurst and J. L. Willis, Class. Quant. Grav. {\bf 20}, 1031 (2003).
\bibitem{ashtekar_2}
G. M. Hossain, V. Husain and S. S. Seahra, Class. Quant. Grav. {\bf 27}, 165013 (2010).

\bibitem{veneziano_1}
G. Veneziano, Europhys. Lett. {\bf 2}, 199 (1986). 
\bibitem{veneziano_2} 
E. Witten, Phys.~Today {\bf 49N4}, 24-30 (1996).
\bibitem{veneziano_3}
F. Scardigli, Phys.~Lett.~ {\bf B452}, 39 (1999).
\bibitem{veneziano_4}
D.  J. Gross and P. F. Mende, Nuc. Phys. {\bf B303}, 407 (1988).
\bibitem{veneziano_5}
D. Amati, M. Ciafaloni and G. Veneziano, Phys, Lett. {\bf B216}, 41 (1989).
\bibitem{veneziano_6}
 T. Yoneya, Mod. Phys. Lett. {\bf A4}, 1587 (1989)

\bibitem{majid} 
S. Majid, 
J. Phys. Conf. Ser. 284, 012003 (2011).

\bibitem{nicolini} M. Isi, J. Mureika and P. Nicolini, JHEP~{\bf 1311}, 139 (2013).

\bibitem{maggiore_1}
M. Maggiore, Phys. Lett. {\bf B 304}, 65 (1993).
\bibitem{maggiore_2}
M. Maggiore, Phys. Lett. {\bf B 319}, 83 (1993).
\bibitem{maggiore_3}
M. Maggiore,  Phys. Rev. {\bf D 49}, 5182 (1994).

\bibitem{carr_1}
B. J. Carr, Mod. Phys. Lett. {\bf A 28}, 1340011 (2013).
\bibitem{cmp}
B. J. Carr, L. Modesto and I. Pr\'emont-Schwarz, 
arXiv: 1107.0708 (2011). 
\bibitem{carr_2}
B. J. Carr,  {\it The Black Hole Uncertainty Principle Correspondence}, to appear in Proceedings of Schwarzschild meeting (Frankfurt July 2013), arXiv1402.1427  (2014).

\bibitem{casadio_1}
R. Casadio, \textit{What is the Schwarzschild radius of a quantum mechanical particle?}, to appear in Proceedings of Schwarzschild meeting (Frankfurt July 2013), arXiv:1310.5452 (2013). 

\bibitem{casadio_2}
R. Casadio and F. Scardigli, Eur.\ Phys.\ J.\ C {\bf 74}, 2685 (2014).

\bibitem{poly_1}
  L. Modesto, 
   Phys. Rev. {\bf D 70}, 124009 (2004).

\bibitem{poly_2}
  L. Modesto, 
  Class. Quant. Grav. {\bf 23}, 5587-5602 (2006).
\bibitem{poly_3}
  L. Modesto, 
  Adv.~High~Energy.~Phys. {\bf 2008}, 459290 (2008).
\bibitem{poly_4}
  L. Modesto, 
   Int.~J.~Theor.~Phys.~{\bf 49},1649-1683 (2010).
\bibitem{poly_6}
  L. Modesto,   arXiv:0811.2196  (2008).
\bibitem{poly_7}
L. Modesto and I. Premont-Schwarz,
Phys. Rev. \textbf{D 80}, 064041 (2009).

\bibitem{bonnano} A. Bonanno and M. Reuter, Phys. Rev.~{\bf D 73}, 083005 (2006).

\bibitem{scardigli} P. Jizba, H. Kleinert and F. Scardigli, AIP Conf. Proc.~{\bf 1446}, 181-189 (2012).

\bibitem{carroll} S. Carroll, \textit{Spacetime and Geometry}, Addison Wesley P.C., San Francisco 2004. 

\bibitem{alfio} 
  A.~Bonanno and M.~Reuter,
  Phys.\ Rev.\ D {\bf 62}, 043008 (2000).

\bibitem{piero} 
  P.~Nicolini, A.~Smailagic and E.~Spallucci,
  Phys.\ Lett.\ B {\bf 632}, 547 (2006).
  
\bibitem{piero2} 
  P.~Nicolini,
  Int.\ J.\ Mod.\ Phys.\ A {\bf 24}, 1229 (2009).
  
\bibitem{pnes10} 
  P.~Nicolini and E.~Spallucci,
  Class.\ Quant.\ Grav.\  {\bf 27}, 015010 (2010).

  
\bibitem{mmn11} 
  L.~Modesto, J.~W.~Moffat and P.~Nicolini,
  Phys.\ Lett.\ B {\bf 695}, 397 (2011).
  
\bibitem{piero3} 
  P.~Nicolini,
  arXiv:1202.2102 (2012).

\bibitem{piero4} 
  P.~Nicolini and E.~Spallucci,
  Adv.\ High Energy Phys.\  {\bf 2014}, 805684 (2014).

\bibitem{dvali} 
  G.~Dvali, G.~F.~Giudice, C.~Gomez and A.~Kehagias,
  JHEP {\bf 1108}, 108 (2011).
  
  
\bibitem{antonio1} 
  A.~Aurilia and E.~Spallucci,
  Adv.\ High Energy Phys.\  {\bf 2013}, 531696 (2013).

\bibitem{antonio2} 
  A.~Aurilia and E.~Spallucci,
  arXiv:1309.7186 (2013).

\bibitem{sabine} 
  S.~Hossenfelder and L.~Smolin,
  Phys.\ Rev.\ D {\bf 81}, 064009 (2010).

\bibitem{robb1} R.~B.~Mann,  A.~Shiekh, L.~Tarasov, Nucl.~Phys.~{\bf B 341}, 134 (1990). 

\bibitem{robb2} 
  A.~E.~Sikkema and R.~B.~Mann,
  Class.\ Quant.\ Grav.\  {\bf 8}, 219 (1991).

\bibitem{robb3} 
  S.~M.~Morsink and R.~B.~Mann,
  Class.\ Quant.\ Grav.\  {\bf 8}, 2257 (1991).
  
\bibitem{robb4} 
  R.~B.~Mann, S.~M.~Morsink, A.~E.~Sikkema and T.~G.~Steele,
  Phys.\ Rev.\ D {\bf 43}, 3948 (1991).
  
  


\bibitem{robb5} 
  R.~B.~Mann and T.~G.~Steele,
  Class.\ Quant.\ Grav.\  {\bf 9}, 475 (1992).


\bibitem{robb6} R.~B.~Mann, Nucl.~Phys.~{\bf B418}, 231 (1994).

\bibitem{robb7} R.~B.~Mann and S.~F.~Ross, Class.~Quant.~Grav.~{\bf 10}, 1405 (1993).

\bibitem{jack1} 
  D.~Cangemi and R.~Jackiw,
  Phys.\ Rev.\ Lett.\  {\bf 69}, 233 (1992).
  
\bibitem{jack2} 
  R.~Jackiw,
  Theor.\ Math.\ Phys.\  {\bf 148}, 941 (2006)
  [Teor.\ Mat.\ Fiz.\  {\bf 148}, 80 (2006)].

  
\bibitem{jack3} 
  D.~Grumiller and R.~Jackiw,
  \textit{Liouville gravity from Einstein gravity},
  Recent developments in theoretical physics,  S. Gosh, G. Kar, 2010. World Scientific, Singapore,2010, p.331, arXiv:0712.3775 (2007).

\bibitem{grum1} 
  D.~Grumiller, W.~Kummer and D.~V.~Vassilevich,
  Phys.\ Rept.\  {\bf 369}, 327 (2002).

\bibitem{grum2} 
  D.~Grumiller and R.~Meyer,
  Turk.\ J.\ Phys.\  {\bf 30}, 349 (2006).

\bibitem{jmpnprd} 
  J.~R.~Mureika and P.~Nicolini,
  Phys.\ Rev.\ D {\bf 84}, 044020 (2011).

\bibitem{thooft} 
  G.~'t Hooft, \textit{Dimensional reduction in quantum gravity},
  Salamfest 1993:0284-296, [gr-qc/9310026] (2013).



\bibitem{cdt1}  R.~Loll, Nucl.~Phys.~Proc.~Suppl.~{\bf 94}, 96 (2001).

\bibitem{cdt2}  
J.~Ambjorn, J.~Jurkiewicz,~R. Loll, Phys.~Rev.~{\bf D 72}, 064014 (2005).

\bibitem{cdt3} 
  J.~Ambjorn, J.~Jurkiewicz and R.~Loll,
  Phys.\ Rev.\ Lett.\  {\bf 95}, 171301 (2005).
  
\bibitem{modestocqg} 
  L.~Modesto,
  Class.\ Quant.\ Grav.\  {\bf 26}, 242002 (2009).
\bibitem{dario} 
  D.~Benedetti,
  Phys.\ Rev.\ Lett.\  {\bf 102}, 111303 (2009).


\bibitem{lmpn} L.~Modesto and P.~Nicolini, Phys.~Rev.~{\bf D 81}, 104040 (2010). 

\bibitem{pnes}
  P.~Nicolini and E.~Spallucci,
  Phys.\ Lett.\ B {\bf 695} (2011) 290.

\bibitem{Carlip2009}  S.~Carlip,  \textit{The Small Scale Structure of Spacetime}, in  Foundations of Space and Time, George Ellis, Jeff Murugan, Amanda Weltman (eds.), Cambridge University Press (2011).

\bibitem{calcagni} 
  G.~Calcagni,
  Phys.\ Rev.\ Lett.\  {\bf 104}, 251301 (2010).



\bibitem{ds} D.~Stojkovic,   Rom.\ J.\ Phys.\  {\bf 57}, 992 (2012).
\bibitem{jrmds1} J.~R.~Mureika and D.~Stojkovic, Phys.~Rev.~Lett.~{\bf 106}, 101101 (2011).
\bibitem{jrmds2}J.~R.~Mureika and D.~Stojkovic Phys.~Rev.~Lett.~{\bf 107}, 169002 (2011). 


\bibitem{rinaldi2} M.~Rinaldi, Class.~Quant.~Grav.~{\bf 29}, 085010 (2012) 085010. 

\bibitem{jrmbh} J.~R.~Mureika, Phys.~Lett.~{\bf B 716}, 171-175 (2012). 


\bibitem{jmpnepj} J.~Mureika and P.~Nicolini, Eur.~Phys.~J.~Plus~{\bf 128}, 78 (2013). 


\bibitem{rainbow} J.~Magueijo, L.~Smolin, Class.~Quant.~Grav.~{\bf 21}, 1725-1736 (2004). 



\bibitem{camacho} A.~ Camacho, Gen.~Rel.~Grav.~{\bf 35} (2003) 319-325. 


\bibitem{holzhey1} 
  C.~F.~E.~Holzhey and F.~Wilczek,
  Nucl.\ Phys.\ B {\bf 380}, 447 (1992).

\bibitem{holzhey2} 
C.~Sivaram and K.~P.~Sinha,
  Phys.\ Rev.\ D {\bf 16}, 1975 (1977).

\bibitem{holzhey3} 
    A.~Salam and J.~A.~Strathdee,
  Phys.\ Rev.\ D {\bf 18}, 4596 (1978).

\bibitem{holzhey4} 
 R.~L.~Oldershaw,
  J.\ Cosmol.\  {\bf 6}, 1361 (2010).

 
 
 
\bibitem{hawking1} S. W. Hawking, Nature \textbf{248}, 30 (1974).

\bibitem{hawking2} S. W. Hawking, Comm. Math. Phys. \textbf{43}, 199 (1975).

\bibitem{vafa} 
  A.~Strominger and C.~Vafa,
  Phys.\ Lett.\ B {\bf 379}, 99 (1996).

\bibitem{carlo} 
  C.~Rovelli,
  Phys.\ Rev.\ Lett.\  {\bf 77}, 3288 (1996).
  
\bibitem{robb98} 
  R.~B.~Mann and S.~N.~Solodukhin,
  Nucl.\ Phys.\ B {\bf 523}, 293 (1998).

\bibitem{kaul} 
  R.~K.~Kaul and P.~Majumdar,
  Phys.\ Rev.\ Lett.\  {\bf 84}, 5255 (2000).


\bibitem{rbmjm} J.~R.~Mureika and R.~B.~Mann, Mod.~Phys.~Lett.~{\bf A26}, 171-181 (2011). 

\bibitem{ch1974}
B.~J.~Carr and S.~W.~Hawking, Mon.Not.Roy.Astron.Soc. {\bf 168}, 399 (1974).



\bibitem{carr3} B.~J.~Carr, K.~Kohri, Y.~Sendouda, J.~Yokoyama, Phys.~Rev.~{\bf D 81}, 104019 (2010). 
\bibitem{nieto} A.~S~.Goldhaber, M.~M.~ Nieto, Rev.~Mod.~Phys.~{\bf 82}, 939-979 (2010). 

\bibitem{lake}
M.~J.~Lake and B.~J.~Carr, ``The Compton-Schwarzschild correspondence from extended de Broglie relations'', arXiv:1505.06994 [gr-qc].
\end{thebibliography}
\end{document}